
{\leftskip=0pt plus4em \rightskip=\leftskip
\parfillskip=0pt \spaceskip=.3333em \xspaceskip=.5em
\pretolerance=9999 \tolerance=9999
\hyphenpenalty=9999 \exhyphenpenalty=9999 }


\magnification = 1200
\baselineskip 14pt plus 2pt
\bigskip

\centerline{\bf THE CLOUD-IN-CLOUD PROBLEM IN THE PRESS-SCHECHTER FORMALISM}
\centerline{\bf OF HIERARCHICAL STRUCTURE FORMATION}
\vskip 0.7in
\centerline{Karsten Jedamzik}
\centerline{Physics Research Program}
\centerline{Institute for Geophysics and Planetary Physics}
\centerline{University of California}
\centerline{Lawrence Livermore National Laboratory}
\centerline{Livermore, CA 94550}
\vskip 1.35in

\centerline{\bf Abstract}
The formalism by Press and Schechter (PS) is often used to infer number
densities of virialized objects of mass $M$ (e.g. quasars,
galaxies, clusters of galaxies, etc.) from a count of
initially overdense regions in
a Gaussian density perturbation field. We reanalyze the PS-formalism by
explicitly counting underdense regions which are embedded within overdense
regions, so called cloud-in-clouds. In contrast to the original PS-formalism,
our
revised analysis automatically accounts for all the cosmic material. We find
that mass
distribution functions for virialized objects are altered by
the proper solution of the cloud-in-cloud problem. These altered distribution
functions agree much better with distribution functions
inferred from N-body simulations than the original PS-distribution functions.

\vskip 1.35in
\centerline{\it Subject headings: cosmology - theory, large-scale structure
of the universe}
\vfill\eject

\centerline{\bf 1. Introduction}

In our current understanding the presently observed structure of the universe
has formed by the gravitational growth of small-amplitude density perturbations
extant at the epoch of matter-radiation equality at high redshift ($z\approx
1000$). It is well known that a detailed comparison between quantities of the
observed structure (e.g. spatial and angular correlation functions, the
peculiar
velocity field, the number densities of quasars, damped Lyman-$\alpha$ clouds,
and galaxy clusters) and predictions of these quantities in specific galaxy
formation scenarios can yield valuable information about the matter content of
the universe and the nature of the initial density perturbations. For a
reliable
prediction of the structure formed in specific cosmic scenarios one has, in
principle, to resort to detailed N-body simulations. In practice, however, such
N-body simulations may be rather time-consuming and can currently only resolve
a
limited dynamic range of mass.

A complimentary and analytic approach to the problem of structure formation has
been developed by Press \& Schechter 1974 (hereafter; PS). In their analysis
the
number densities and masses of virialized objects is directly inferred from the
statistics of the initial density perturbation field in a manner to be
illustrated in the following section. The PS-prediction for the fraction of
cosmic matter contained in virialized and separate halos of mass $M$, so called
multiplicity- or mass- functions, is in fairly good agreement with results from
N-body calculations indicating the general validity of the PS-approach. The
PS-mass functions have been applied to predict the number densities of galaxies
(Cole \& Kaiser 1989; White \& Frenk 1991; Kauffmann, White, \& Guiderdoni
1993),
clusters of galaxies (Cole \& Kaiser 1988; Zhan 1990; Bartlett \& Silk 1993),
quasars (Carlberg 1990; Nusser \& Silk 1993),
Lyman-$\alpha$ clouds (Kauffmann \& Charlot 1994; Mo \& Miralda-Escud\'e 1994),
and dark halos (Narayan \& White 1988)
which emerge in different structure formation scenarios and at different
redshifts. Recently the analysis by PS has also been extended to estimate the
frequency of halo mergers and to predict details about the substructure of
clusters of halos (Bower 1991; Bond {\it et al.} 1991; Kauffmann \& White 1993;
Lacey \& Cole 1993; 1994).

It has been noted, however, that the original analysis by PS
is strictly speaking incorrect since it only associates half of the mass of the
initial density perturbation field with eventually to be virialized and
gravitationally self-bound structures.
PS achieved proper normalization of the mass functions by simply multiplying
results by a factor of two.
The remaining half of cosmic material, which is not automatically accounted for
in the PS-formalism, is material initially present in underdense regions.
It is believed that this remaining half of material is accounted for
if a proper solution to the cloud-in-cloud problem is found, in particular the
problem of correctly counting underdense regions which are embedded within
overdense regions. The cloud-in-cloud problem has also plagued other analytic
approaches to the subject of structure formation, such as the peaks formalism
(Peacock \& Heavens 1985; Bardeen {\it et al.} 1986).

Bond {\it et al.} (1991) proposed a solution to the cloud-in-cloud problem by
considering excursion set mass functions. They were able to obtain analytic
results for the mass functions if they applied sharp $k$-space filtering.
In sharp $k$-space filtering average densities within a given volume
are computed only from those Fourier modes of the density distribution which
have wavelength larger than the characteristic dimension of the volume
(refer to the next section).
Their analysis accounts for all the cosmic material.
Surprisingly, their analysis recovers the original PS-mass functions
{\it including} the normalization factor of two. A
similar result has been obtained by Peacock and Heavens (1990).
When different filtering functions are considered (e.g. top hat or Gaussian
filters in coordinate space) the shapes of mass functions may be altered.

In this paper we reanalyze the cloud-in-cloud problem in the context of the
PS-formalism. The method outlined is an extension of the original PS-analysis
which reduces to the PS-prescription when the cloud-in-cloud problem is
(incorrectly) ignored. We find that a proper treatment
of the cloud-in-cloud problem
changes the shapes of mass functions, in contrast to the conclusions of
Bond {\it et al.} (1991) and Peacock \& Heavens (1990).
We show that the renormalized mass functions are in much better agreement with
results from N-body simulations than the original PS mass functions.

\vskip 0.15in
\centerline{\bf 2. The cloud-in-cloud problem and mass distribution functions}
\vskip 0.15in

It is well known that length and amplitude of small-amplitude density
perturbations (i.e. amplitude $\delta\equiv (\rho-{\bar\rho})/{\bar\rho}<<1$;
where $\rho$ is the mass density of the perturbation and $\bar\rho$ is the
average cosmic mass density) grow proportionally to the cosmic scale factor
during a matter-dominated epoch. When reaching non-linearity ($\delta\sim 1$)
perturbations recollapse and virialize. (cf. Kolb \& Turner 1989). In the
absence of dissipative processes
fluctuations will maintain their sizes and densities thereafter.
Perturbations which are overdense by more than some critical amount
$\delta_c(z)$ at the epoch of matter-radiation density equality
will have recollapsed before the epoch with redshift $z$.
The critical amplitude $\delta_c$ is approximately $\delta_c\approx
1.68(1+z)/(1+z_{eq})$ (cf. Kolb \& Turner 1989).
Here $z_{eq}$ is the redshift of matter-radiation density equality.

One can probe the initial density perturbation field by determining the
fraction of space which is overdense by the critical amount $\delta_c$ when
averaged over a spherical volume $V$ (or, equivalently, on a mass scale
$M={\bar\rho}V$). Presently most structure formation scenarios assume the
existence of Gaussian density perturbations as the seeds for the cosmic
structure.
In this case, the probability $f(\delta_c,M)$ to
find a region of mass scale $M$
to be overdense by more than the critical amount $\delta_c$ is simply given by
an integral over the tail of a Gaussian distribution function
$$f(\delta_c,M)=\int\limits_{\delta_c}^{\infty}{1\over
{\sqrt{2\pi}}}{1\over\sigma (M)}{\rm exp}\biggl(-{1\over
2}{\delta^2\over\sigma^2(M)}\biggr)d\delta\ .\eqno(1)$$

The width of this Gaussian distribution function, $\sigma (M)$, can be inferred
from the statistics of the Fourier modes of the density distribution.
For Gaussian random fluctuations the magnitude of the amplitudes of the
Fourier modes $|\delta_k|$ are distributed according to a half-Gaussian.
The $\phi_k$ follow a flat distribution in the interval
$\phi_k\in [0,2\pi]$ so that different Fourier modes have uncorrelated
phases $\phi_k$.
In this case, the overdensity
$\delta (x)=2\sum_k|\delta_k|{\rm cos}(kx+\phi_k)$ at some coordinate $x$
is the sum over several uncorrelated terms. By the central limit theorem
the distribution function for $\delta$ will then be a Gaussian centered
around value zero.

If one wishes to determine the average density
$\delta_M(x)$ on a mass scale $M$ (or in a volume $V=M/{\bar\rho}$) only
those Fourier modes should be added which have wavelengths $(2\pi /k)$ larger
than the characteristic dimension of the volume.
Such a prescription corresponds to sharp $k$-space filtering.
It is expected that the
contributions from Fourier modes which have wavelengths shorter than the
characteristic
dimension of the volume will average out to zero.

The variance $\sigma (M)$ of the resultant distribution
function for $\delta_M$ can be
computed from the variance $\langle|\delta_k|^2\rangle$ of the
half-Gaussian distribution function for $|\delta_k|$.
Here the square brackets $\langle \rangle$ denote an ensemble average.
The ensemble includes different universes with different randomly chosen
$\delta_k$ and $\phi_k$ distributed according to the half-Gaussian- and
flat distribution functions, respectively.
Note, that we assume implicitly that an ensemble average is
equivalent to a spatial average. Then
$$\sigma^2(M)\equiv\langle\delta_M^2\rangle\sim {L^3\over (2\pi )^3}
\int_0^{k_c}dk\ k^2\langle|\delta_k|^2\rangle\ .\eqno(2a)$$
In this expression $L^3k^2dk/(2\pi )^3$ counts the number of different Fourier
modes
and $L$ is the length of the fundamental cube of the simulation.
(Results will emerge independent from $L$.)
Bond {\it et al.} (1991) have noted that equation (2a) is completely analogous
to computing the mean-square-distance $\langle d^2\rangle$
(corresponding to $\sigma^2(M)$) which a particle random walking in one
dimension will have moved away from it's initial position, given the
expectation value of the square of the particle's mean free path
$\langle l^2\rangle$ (corresponding to $\langle|\delta_k|^2\rangle$)
and the number of scatterings $N$ (corresponding to the number of
Fourier modes). In particular $\langle d^2\rangle =N\langle l^2\rangle$.

The sharp $k$-space cutoff $k_c$ in equation (2a) is determined by the mass of
the probing volume $M$
$$k_c(M)\sim V^{-{1\over 3}}\sim \biggl({M\over{\bar\rho}}\biggr)^{-{1\over
3}}\ .\eqno(2b)$$
The relationship between the cutoff wavenumber $k_c$ and the mass $M$ is
ambiguous up to factors of unity. This is because it is not evident if the
diameter
of a spherical volume $V$ should be set equal to either the cutoff wavelength
or, for example, half the cutoff wavelength. This ambiguity results in an
uncertainty of what mass should ultimately be associated with an initially
overdense region. It does, however, not affect the shape of the resulting mass
function. Others have chosen a value of $k_c^3=6\pi^2(M/{\bar\rho})^{-1}$
(Lacey \& Cole 1994).

One often assumes a featureless power-law for the variance of the magnitude
of the Fourier amplitude, i.e. $\langle|\delta_k|^2\rangle=L^{-3}Ak^n$.
In this case, equation (2a) yields
$$\sigma^2(M)=\biggl({M\over M_{\star}}\biggr)^{-(1+{n\over 3})}.\eqno(3)$$
Here $M_{\star}$ is the mass scale of unity variance $\sigma^2(M_{\star})=1$.
The coefficient $A$ and the exact relationship between mass and cutoff momentum
$k_c(M)$ are included in $M_{\star}$.

In specific scenarios of structure formation and
if one desires to work over extended ranges of mass
the featureless power-law should be replaced by the appropriate functional form
for $\langle|\delta_k|^2\rangle$ which, in general, is more complicated.
The dependence of $\langle|\delta_k|^2\rangle$ on $k$ determined at the epoch
of matter-radiation density equality for a variety of cosmic scenarios can, for
example, be taken from the paper by Holtzmann (1989).
Equation (2a) can then be numerically integrated to yield an analogous
relationship to
equation (3).

It is now our goal to compute the number density of {\it isolated} regions with
an overdensity $\delta$ exceeding the critical one, $\delta\geq\delta_c$.
Isolated regions overdense by $\delta_c$ or more are regions which are not
included in yet larger regions overdense by $\delta_c$ or more. In particular,
a
region of mass scale $M$ will only be counted as an eventually virialized
object
of mass $M$, if for any larger mass scale the average overdensity
of the region is below the
threshold. Consider a set of mass scales $(M_1,M_2,M_3,...)$ which are ordered
by size $(M_1<M_2<M_3<...)$. Furthermore, consider a large representive cosmic
volume $L^3$. The total volume, $L^3f(\delta_c,M_1)$, which is found to be
overdense by $\delta_c$ or more on the smallest scale $M_1$ is given by the
number of isolated objects of scale $M_1$, $L^3n^{\star}(M_1)$, times the size
of an individual object's volume,
$V_1=M_1/{\bar\rho}$, plus the contributions from finding
overdense regions on scale $M_1$ included in larger isolated regions
$(M_2,M_3,...)$. The latter contribution, for example from isolated regions on
scale $M_2$, is given by the product of number of isolated regions on scale
$M_2$, $L^3n^{\star}(M_2)$, the volume of these isolated regions,
$V_2=M_2/{\bar\rho}$, and the probability $P(M_1,M_2)$ of finding a region of
scale $M_1$ overdense by $\delta_c$ or more, provided it is included in an
isolated, overdense region of size $M_2$. In the original PS-analysis it was
erroneously assumed that this probability is unity. In a similar manner
isolated
regions of even larger sizes make contributions to $L^3f(\delta_c,M)$.

When we add up all contributions we find
$$f(\delta_c,M_1)={1\over{\bar\rho}}\sum_{i=1}^{\infty}M_in^{\star}(M_i)P(M_1,M_i)\ ,\eqno(4a)$$
where the limiting probability $\lim_{M_1\to M_2} P(M_1,M_2)$
is understood to be unity. Equation (4a) can be rewritten into integral form
$$f(\delta_c,M_1)={1\over{\bar\rho}}\int_{M_1}^{\infty}dM^{\prime}\
M^{\prime}n(M^{\prime})P(M_1,M^{\prime})\ .\eqno(4b)$$
Here $n(M)$ is the number density of isolated and overdense regions per unit
mass interval. This is exactly the desired mass function which PS
intended to derive.
Completely
analogous equations to equations (4ab) can be written down for any mass scale
$M_j$. The
resulting set of equations then represent a matrix equation, such that the
product of matrix $P(M_j,M_i)$ and vector $M_in^{\star}(M_i)$ has to equal the
vector $f(\delta_c,M_j)$. This matrix has a form which is already suitable for
Gaussian back
substitution.
This is because the $n(M_j)$ only depend on those $n(M_i)$ with $M_i>M_j$ and
so
all lower off-diagonal elements of the matrix
$P_{ji}\equiv P(M_j,M_i)$ are zero (i.e. $P_{ji}=0$, for $j>i$).
To obtain $n(M)$ one has,
in principle, to solve an infinitely large matrix equation.
In practice, however, one can obtain very good approximations to
$n(M)$ by truncating the matrix for large masses. The number density of
isolated, overdense regions with large masses is exponentially suppressed so
that the upper, infinite bound in equations (4ab) can be replaced by some large
mass
$M_{trunc}$.

To fully appreciate the modification to the original PS-formalism one can take
the derivative of equation (4b) with respect to mass $M_1$. This yields
$${df(\delta_c,M)\over dM}=-{1\over{\bar\rho}}Mn(M)+
{1\over{\bar\rho}}\int_M^{\infty}dM^{\prime}M^{\prime}n(M^{\prime}){dP(M,M^{\prime})\over dM}\ .\eqno(5)$$
Note, that by neglecting the second term on the right-hand side of equation
(5), the
original PS-prescription is recovered. However, this term does not vanish since
the
probability $P(M,M^{\prime})$ does vary with $M$.

The probability $P(M,M^{\prime})$ to find a region of size $M$ overdense by
$\delta_c$ or more when it is included in a larger region $M^{\prime}$
overdense
by $\delta_c$ or more can be computed in the following way. The overdensity
$\delta^{\prime}$ of the large region is determined by the sum of the
amplitudes
of all the Fourier modes between $k=0$ and the cutoff $k=k_c^{\prime}$, where
the cutoff is computed from equation (2b) for the mass $M^{\prime}$.
Substructure
within the large region, that is the overdensity (underdensity) for any smaller
region $M$ contained within $M^{\prime}$ {\it relative} to the overdensity
$\delta^{\prime}$, is determined by the sum of the amplitudes of all the
Fourier
modes which have wave numbers between $k_c^{\prime}$ and $k_c$. Here $k_c$ is
computed for the smaller mass $M$. The probability distribution for the
overdensity $\delta$ of the smaller region will follow a Gaussian distribution
centered at the larger region's overdensity $\delta^{\prime}$. If one
assumes a power-law for the variances $\langle|\delta_k|^2\rangle$ the width
$\sigma_{sub}$ of this Gaussian can be obtained from equation (2a).
This is accomplished if one replaces
the lower integral limit in equation (2a) by $k_c^{\prime}$
$$\sigma^2_{sub}=\biggl({M\over M_{\star}}\biggr)^{-(1+{n\over
3})}-\biggl({M^{\prime}\over M_{\star}}\biggr)^{-(1+{n\over 3})}.\eqno(6)$$

The probability $P(M,M^{\prime})$ is then computed by a double integral.
The first integral is performed over all possible larger region's
overdensities having $\delta^{\prime}\geq \delta_c$. The second integral
computes for any given $\delta^{\prime}$ the probability of finding a smaller
region's overdensity to exceed the threshold $\delta\geq\delta_c$ as well.
This latter integral is obtained by integrating over the Gaussian
distribution of the smaller region's overdensity. We find
$$P(M,M^{\prime})=N^{-1}\int_{\delta_c}^{\infty}d\delta^{\prime}{1\over\sqrt{2\pi}}{1\over\sigma^{\prime}}{\rm exp}\biggl(-{1\over 2}{{\delta^{\prime}}^2\over{\sigma^{\prime}}^2}\biggr) \int_{\delta_c}^{\infty}d\delta {1\over\sqrt{2\pi}}{1\over\sigma_{sub}}{\rm exp} \biggl(-{1\over 2}{(\delta-\delta^{\prime})^2\over\sigma_{sub}^2}\biggr), \eqno(7a)$$
where the constant $N$ assures proper normalization
$$N=\int_{\delta_c}^{\infty}d\delta^{\prime}{1\over\sqrt{2\pi}}{1\over
\sigma^{\prime}}{\rm exp}\biggl(-{1\over
2}{{\delta^{\prime}}^2\over{\sigma^{\prime}}^2} \biggr).\eqno(7b)$$
The width $\sigma^{\prime}$ of the larger region's overdensity
distribution is given by equation (3) with $M$ replaced by $M^{\prime}$. In the
limit
$M\to M^{\prime}$ the width $\sigma_{sub}$ approaches zero and the second
integral in eq(7a) becomes a delta function. This yields a limiting
probability of
$\lim_{M\to M^{\prime}} P(M,M^{\prime})=1$. In the opposite limit when $M$
approaches zero the probability $P(M,M^{\prime})$ approaches $1/2$.

It is the goal to solve the implicit equation (5) for $n(M)$. We have derived a
rather lengthy expression for $dP(M,M^{\prime})/dM$ which, however, is more
suitable for numerical integration than equation (7a)
$${dP(M,M^{\prime})\over dM}=\Biggl({\rm erfc}\biggl({\delta_c\over\sqrt 2
\sigma^{\prime}}\biggr)\Biggr)^{-1}\times\Biggl({1\over\pi}
{1\over\sigma^{\prime}}{1\over\sigma_{sub}^2}{d\sigma_{sub}\over dM}
{\rm exp}\biggl(-{1\over 2}{\delta_c^2\over
(\sigma_{sub}^2+{\sigma^{\prime}}^2)}\biggr)\Biggr)\times$$
$$\times\Biggl(\delta_c \biggl({\pi\over 2}\biggr)^{1\over 2}
{\sigma_{sub}^3\sigma^{\prime}\over (\sigma_{sub}^2+{\sigma^{\prime}}^2)
^{3\over 2}}{\rm erfc}\biggl({\delta_c\over\sqrt 2}
{\sigma_{sub}\over\sigma^{\prime}}{1\over (\sigma_{sub}^2+{\sigma^{\prime}}^2)
^{1\over 2}}\biggr)\eqno(8)$$
$$-{\sigma_{sub}^2{\sigma^{\prime}}^2\over
(\sigma_{sub}^2+{\sigma^{\prime}}^2)}
{\rm exp}\biggl(-{1\over 2}{\sigma_{sub}^2\over{\sigma^{\prime}}^2}
{\delta_c^2\over (\sigma_{sub}^2+{\sigma^{\prime}}^2)}\biggr)\Biggr)\ .$$
Here erfc($x$) is the complimentary error function
$${\rm erfc}(x)={2\over\sqrt{\pi}}\int_x^{\infty}{\rm e}^{-t^2}dt\ .$$
The derivative $d\sigma_{sub}/dM$ can be obtained from equation (6).

It is now straightforward to solve equation (5) for $n(M)$ with the help of
equation (1) and
equation (8). This is done by assuming a very large mass $M_{trunc}$ for which
$df(\delta_c,M)/dM$ is exponentially small.
For such a large mass one obtains a
good approximation to $n(M_{trunc})$ by only considering the first term
on the right-hand-side of equation (5).
To compute $n(M)$ for smaller masses the full equation (5) should be employed.
This is easily done if one derives $n(M)$ for successively smaller masses by
using previously determined $n(M)$ for all the larger masses. In this way the
shape of the original PS-mass function will be altered.

We have numerically solved equation (5) to obtain the multiplicity function.
(We will give an analytic approximation below.)
The multiplicity
function is the fraction of matter $d\rho (M)/{\bar\rho}$
contained in virialized objects
of mass $M$ per unit logarithmic mass interval $d{\rm log}_2M$. The
multiplicity
function is related to $n(M)$ by
$${1\over{\bar\rho}}{d\rho (M)\over d{\rm log}_2M}=
{1\over{\bar\rho}}M^2n(M){\rm log_e}2\ .\eqno(9)$$

Results for multiplicity functions are presented in figure (1).
In this figure, the upper panel shows the multiplicity function for a
spectral index $n=+1$, whereas the lower panel shows the multiplicity
function for a spectral index of $n=-1$.
The heavy solid lines show the renormalized multiplicity
functions derived in this paper. For comparison the light solid lines represent
the original PS-results. The
figure also shows multiplicity functions inferred from N-body simulations.
These have been taken from Efstathiou {\it et al.} (1988) (cf. figs. 9 in
Efstathiou {\it et al.}).
The multiplicity functions inferred from N-body simulations are shown for the
last seven output times of the simulation and are scaled according to
self-similarity of the multiplicity
functions at different times. The up-turn of the multiplicity functions for
lower multiplicities or, equivalently, masses arises from the limited
resolution of the N-body simulations.

The figure illustrates that the multipicity functions derived from equation
(5), which include
the solution of the cloud-in-cloud problem, provide a much better approximation
of the N-body results than the original PS-multiplicity functions.
Over the limited range of masses (multiplicities) shown in figure (1) the
differences between the PS-result and the result from equation (5) are within a
factor of two.

In figures (2abc) we show the multiplicity functions on a double-logarithmic
plot over an extended mass range. The solid lines represent the renormalized
multiplicity functions, whereas the dashed lines show the original PS
multiplicity functions. It is evident that the differences between these
two multiplicity functions can be as large as an order of magnitude at the
small-mass scale side of the distribution. In particular, the PS-analysis
systematically underestimates the number densities of small mass objects.
This discrepancy between our result and the PS-approximation is particularly
large for a positive spectral index. On the large-mass scale end of the
distribution the PS multiplicity functions overestimate the number densities of
objects by a factor of two independent of the spectral index. However,
this factor-of-two difference can be easily absorbed into the uncertainties
associated with the absolute mass scale of objects (cf. equation 2b).
This is because there is no change in the shape
of the multiplicity functions at the rare, massive-object side of the
distribution
and multiplicity functions decrease very rapidly for large masses.

The modification of the mass distribution functions presented here does not
affect the
self-similarity of mass distribution functions at different times
in hierarchical structure formation. In particular,
as described by PS the shapes of the mass
functions are independent of epoch or, equivalently, threshold $\delta_c$. Mass
functions differ only for different $\delta_c$
by an increase in the characteristic mass scale
of the distribution functions with decreasing threshold $\delta_c$.

The formalism developed naturally assigns all the cosmic material
to overdense regions which will eventually form virialized
objects. Thus, the missing half of cosmic material, which could not be
accounted for in the PS-formalism, is accounted for once a solution to
the cloud-in-cloud problem is known.
This can be seen by considering a mass scale $M$ very much smaller than the
characteristic mass scale of the peak of the multiplicity function.
In the limit when $M$ approaches zero both
$f(\delta_c,M)$ and $P(M,M^{\prime})$ approach one-half. With the help of
equation (4b) one can then infer
$$\lim_{M\to
0}{1\over{\bar\rho}}\int_0^{\infty}dM^{\prime}M^{\prime}n(M^{\prime})=1\
,\eqno(10)$$
so that the computed $n(M)$ will automatically have the proper normalization.

Our results are in direct contradiction with the results obtained by Bond {\it
et al.} (1991) and Peacock \& Heavens (1990). Their analysis confirmed the
original PS-result. Bond {\it et al.} consider a set of different-sized
mass scales. For each point in space they determine the largest of these mass
scales $M$ for which the region around the point is overdense by the
critical amount. They then associate such a region with an eventually
self-bound object of mass $M$. They perform such an analysis independently
for each space point by considering all possible realizations of Fourier-mode
amplitudes.
Their analysis is incorrect, however, since it does not allow for correlations
between neighboring space points. In other words, if their analysis finds an
object
of mass $M$ at point $x$ it will also find an object of mass
$M+dM$ at point $x+dx$ and it will count both as separate objects. In reality,
the object of mass
$M$ at $x$ will have a finite size and will include the space point at
$x+dx$, so that only one object of mass $M$ should be counted.
Note, that our treatment implicitly includes spatial correlations by
the determination of the probability function $P(M,M^{\prime})$.

We have not been able to derive an analytic solution for $n(M)$. It is not
always convenient to have to invert a matrix for a determination of
multiplicity functions and number densities of virialized objects. We therefore
obtained an analytic approximation for
$n(M)$
$$n(M)\approx {1\over 2}n_{\rm PS}(M)\bigl(1-0.68y^{-{1\over 5}}+1.17
y^{-{2\over 5}}\bigr)\ ,\eqno(11a)$$
with
$$y=\delta_c\biggl({M\over M_{\star}}\biggr)^{{1\over 2}+{n\over
6}}.\eqno(11b)$$
Here $n_{\rm PS}(M)$ is the original PS-result, in particular
$$n_{\rm PS}(M)={\sqrt{2\over\pi}}\biggl({1\over 2}+{n\over 6}\biggr)
{\bar\rho\over M^2}y{\rm exp}\biggl(-{y^2\over 2}\biggr)\ .\eqno(12)$$
The multiplicity functions derived by using the analytic approximation
equation (11a) are shown in figures (2abc) by the dotted lines. Over the whole
range
of masses shown in these figures and for all three spectral indices equation
(11a)
approximates the renormalized number densities by better than ten per cent.
If higher accuracies are desired a numerical treatment has to be performed.

\vskip 0.15in
\centerline{\bf 3. Conclusions}
\vskip 0.07in

We have reanalyzed the Press-Schechter formalism which is often used to
estimate number densities of virialized objects
from the statistics of the primordial density perturbations.
Our result addresses the importance of the cloud-in-cloud problem, in
particular the correct count of underdense regions which are embedded within
overdense regions. In contrast to previous work we find that by considering
a proper solution to the cloud-in-cloud problem mass distribution functions
are modified. These renormalized mass distribution functions
provide an excellent approximation of mass distribution functions inferred
from N-body simulations. Our analysis shows that the original PS-formalism
systematically overestimates the number of rare, massive objects by a factor
of two, while the number of low-mass objects are underestimated by up to
an order of magnitude. The approach described here avoids this.

\vfill\eject
\centerline{\bf 4. Acknowledgements}
\vskip 0.07in

The author is grateful to G.J. Mathews for many inspiring discussions. The
author wishes also to acknowledge discussions with S. Charlot and G. Kauffmann.
This work was performed under the auspices of the U.S. Department of Energy
by the Lawrence Livermore National Laboratory under contract number
W-7405-ENG-48 and DoE Nuclear Theory Grant SF-ENG-48.
\vfill\eject

\centerline{\bf 5. References}
\vskip 0.1in

\item{ }Bardeen, J.M., Bond, J.R., Kaiser, N., \& Szalay, A. S., 1986,
ApJ, 304, 15
\item{ }Bartlett, J.G. \& Silk, J. 1993, ApJ (Letters), 407, L45
\item{ }Bond, J.R., Cole, S., Efstathiou, G., \& Kaiser, N. 1991,
ApJ, 379, 440
\item{ }Bower, R.G. 1991, MNRAS, 248, 332
\item{ }Carlberg, R.G. 1990, ApJ, 350, 505
\item{ }Cole, S. \& Kaiser, N. 1989, MNRAS, 237, 1127
\item{ }Efstathiou, G., Frenk, C.S., White, S.D.M., Davis, M. 1988,
MNRAS, 235, 715
\item{ }Holtzman, J.A. 1989, ApJS, 71, 1
\item{ }Kauffmann, G. \& White, S.D.M. 1993, MNRAS, 261, 921
\item{ }Kauffmann, G., White, S.D.M., \& Guiderdoni, B. 1993, MNRAS,
264, 201
\item{ }Kauffmann, G. \& Charlot, S. 1994, ApJ, submitted
\item{ }Kolb, E.W. \& Turner, M.S. 1989, in {\it The Early Universe},
Addison-Wesley, pg 327
\item{ }Lacey, C. \& Cole, S. 1993, MNRAS, 262, 627
\item{ }Lacey, C. \& Cole, S. 1994, MNRAS, submitted
\item{ }Mo, H.J. \& Miralda-Escud\'e, J. 1994, ApJ, submitted
\item{ }Narayan, R. \& White, S.D.M. 1988, MNRAS, 231, 97p
\item{ }Nusser, A. \& Silk, J. 1993, ApJ (Letters), 411, L1
\item{ }Peacock, J.A. \& Heavens, A.F. 1985, MNRAS, 217, 805
\item{ }Peacock, J.A. \& Heavens, A.F. 1990, MNRAS, 243, 133
\item{ }Press, W.H. \& Schechter, P. 1974, ApJ, 187, 425
\item{ }White, S.D.M. \& Frenk, C.S. 1991, ApJ, 379, 25
\item{ }Zhan, Y. 1990, ApJ, 355, 387

\vfill\eject

\centerline{\bf 6. Figure Captions}
\vskip 0.15in

\item{\bf Figure 1}The multiplicity function, i.e. the fraction of
cosmic material contained in virialized and isolated objects of mass $M$ per
unit logarithmic mass interval (cf. equation 9) as a function of
${\rm log}_2M$. The absolute value of mass on the ordinate
is arbitrary. The heavy solid line shows the result derived in this paper.
The light solid line shows the original PS-result.
For comparison the solid (up-turning), dotted, short-dashed, long-dashed,
short-dashed dotted, long-dashed dotted, and short-dashed long-dashed lines
give results from N-body simulations. The N-body results have been taken
from Efstathiou {\it et al.} (1988). The upper panel shows the
multiplicity function for a power-law spectral index of $n=+1$ for the
variance of the Fourier mode amplitudes of the density distribution,
whereas the lower panel shows the multiplicity function for a spectral index
of $n=-1$.

\item{\bf Figure 2a}The multiplicity function $(1/\bar\rho )
(d\rho (M)/d{\rm log_2}M)$ presented double-logarithmically as a function of
rescaled mass $\delta_c^{6/(3+n)}(M/M_{\star})$. The solid line shows the
renormalized multiplicity function, whereas the dashed line shows the original
PS-multiplicity function. The dotted line represents the multiplicity
function computed by using the analytic approximation equation (11).
The calculation assumes a spectral index of $n=+1$.

\item{\bf Figure 2b}Same as figure (2a), but here a spectral index of $n=0$ is
assumed.

\item{\bf Figure 2c}Same as figure (2a), but here a spectral index of $n=-1$ is
assumed.

\end